\documentclass[aps,prd,twocolumn,superscriptaddress]{revtex4-1}

\usepackage{graphicx}
\usepackage{dcolumn}
\usepackage{bm}
\usepackage{color}
\usepackage{hyperref}
\usepackage[english]{babel}
\usepackage{amsmath}
\usepackage{comment}

\newcommand{\Msun}{M_{\odot}}
\newcommand{\ma}{m_{a}}
\newcommand{\ga}{g_{{a}\gamma}}
\newcommand{\gten}{g_{\rm{10}}}

\newcommand{\MeV}{\rm{MeV}}
\newcommand{\km}{\rm{km}}
\newcommand{\tpb}{t_{\rm{pb}}}

\newcommand{\heat}{\rm{heat}}
\newcommand{\cool}{\rm{cool}}
\newcommand{\rmand}{\rm{and}}

\newcommand{\tkt}{\textcolor{black}}

\begin{document}

\preprint{APS/123-QED}

\title{\texorpdfstring{Progenitor Dependence of Neutrino-driven Supernova Explosions\\with the Aid of Heavy Axion-like Particles}{First Line - Second Line}}

\author{Tsurugi Takata}
    \affiliation{Department of Applied Physics, Fukuoka University, 8-19-1 Nanakuma, Jonan-ku, Fukuoka 814-0180, Japan}
\author{Kanji Mori}
    \affiliation{Division of Science, National Astronomical Observatory of Japan, 2-21-1 Osawa, Mitaka, Tokyo 181-8588, Japan}
\author{Ko Nakamura}
    \affiliation{Department of Applied Physics, Fukuoka University, 8-19-1 Nanakuma, Jonan-ku, Fukuoka 814-0180, Japan}
\author{Kei Kotake}
    \affiliation{Department of Applied Physics, Fukuoka University, 8-19-1 Nanakuma, Jonan-ku, Fukuoka 814-0180, Japan}
    \affiliation{Institute for Theoretical Physics, University of Wrocław, 50-204 Wrocław, Poland
}


\begin{abstract}
We perform spherically symmetric simulations of core-collapse \tkt{supernovae} with the aid of heavy axion-like particles (ALPs) which interact with photons and redistribute energy within supernova matter.
We explore a wide ALP parameter space that includes MeV-scale ALP mass $\ma$ and the ALP-photon coupling constant $\ga \sim 10^{-10} \, \rm{GeV}^{-1}$ , employing three  progenitor models with zero-age \tkt{main-sequence} mass of $11.2\,\Msun$, $20.0\,\Msun$, and $25.0\,\Msun$. 
We find a general trend that, given $\ma\lesssim300$\,MeV, heavier ALPs are favorable for the shock wave to be successfully revived, \tkt{aiding the onset of the neutrino-driven explosion.} However, if ALPs are heavier than $\sim400$\,MeV, the explosion is failed or weaker than that for the models with smaller $\ma$, because of an insufficient temperature 
inside the supernova core to produce heavy ALPs.
The maximum temperature in the core depends on the initial progenitor structure. Our simulations indicate that the high-temperature environment \tkt{in} the collapsing core of massive progenitors leads to a significant impact of ALPs on the explodability.

\end{abstract}
\maketitle

\section{Introduction}
\tkt{Axion is} a hypothetical particle that is introduced to solve the strong CP problem \cite{wilczek_problem_1978,weinberg_new_1978}.  Axions can interact with the Standard Model particles such as nucleons, photons, and electrons, and hence they can be produced in a hot plasma in astrophysical objects. Motivated by the recent development in the string theory \cite{svrcek_axions_2006,cicoli_type_2012,arvanitaki_string_2010}, a more general class of \tkt{axion-like} particles (ALPs) \tkt{has been} introduced as a new particle beyond the Standard Model, where the mass and the coupling constants are treated as independent parameters.

ALPs that interact with photons have been searched for through astronomical observations. For example, \tkt{the energy loss and transfer} induced by ALPs can shorten the stellar lifetime, particularly for horizontal branch (HB) stars. Since globular clusters contain many HB stars, their stellar populations have been used to constrain the ALP-photon coupling \cite{raffelt_bounds_1987,ayala_revisiting_2014,carenza_constraints_2020,lucente_constraining_2022}.

ALPs can also affect dynamics of core-collapse supernovae (SNe), which are the most copious astrophysical source of ALPs. If ALPs are heavy and decay radiatively inside the star, they can heat the stellar mantle \cite{sung_new_2019}. As a result, the explosion energy can become higher than the standard SN models. Recently, Refs.~\cite{Caputo_low-energy_2022, Caputo_muonic_2022} pointed out that low-energy SN events such as SN 1054 can be used to severely constrain the ALP-photon interaction strength. In addition, if the decay of ALPs takes place outside the star, 
the $\gamma$-rays produced via the ALPs' decay can be detectable on the Earth \cite{jaeckel_decay_2018}. 
The fact that such $\gamma$-rays accompanied by the neutrino burst were not detected from SN 1987A \cite{chupp_experimental_1989} imposed upper limits on the coupling constant \cite{PhysRevD.39.1020, PhysRevLett.60.1797, PhysRevLett.65.960, giannotti_new_2011,payez_revisiting_2015, diamond_axion-sourced_2023, muller_investigating_2023}.

The neutrino burst from SN 1987A, which lasted for $\sim10$\,s \cite{hirata_observation_1987,bionta_observation_1987,alekseev_possible_1987}, is also a useful probe into the production of new particles in SNe. If the ALP luminosity exceeds the neutrino luminosity, the proto-neutron star (PNS) cooling should be significantly accelerated,  resulting in a shorter duration time of the SN 1987A neutrino burst than observed. This energy-loss argument has been applied to constrain the nature of ALPs \cite{PhysRevLett.60.1797, raffelt_bounds_1988, masso_light_1995, lee_revisiting_2018,lucente_heavy_2020,lella_getting_2024}. However, \tkt{most of these studies of the traditional energy-loss argument were not} based on \tkt{realistic numerical} simulations. Since the PNS cooling is a non-linear process, it is necessary to perform hydrodynamic simulations coupled with the ALP emission to predict the neutrino detection rate quantitatively. For example, SN models  developed in Ref.~\cite{betranhandy_neutrino_2022} show that QCD axions can promote the PNS contraction, contributing to the enhancement of the neutrino mean energy.
Thus, since ALPs can change the SN properties, it is necessary to perform self-consistent neutrino-driven SN simulations that take ALPs into account.

ALPs, if they exist, can leave traces in multi-messenger signals 
\tkt{such as neutrinos and gravitational waves from a nearby SN event}. In order to predict the observable signals, it is \tkt{required} to perform self-consistent simulations that are coupled with the ALP transport, as performed in Refs.~\cite{fischer_probing_2016,fischer_observable_2021,betranhandy_neutrino_2022,mori_shock_2022,mori_multimessenger_2023,foguel_analytic_2023}. In a previous study \cite{mori_shock_2022}, they developed a SN simulation code that takes into account the backreactions of ALP production and decay. 
They employed ALPs with mass $\ma = 50$ -- $800 \,\MeV$ and the ALP-photon coupling constant $\gten = g_{\alpha \gamma} / (10^{-10} \,\rm{GeV^{-1}}) =$ 4 -- 40. As a result, they found that for heavy ALPs with the mass of $\sim100$\,MeV, shock revival can occur even in one-dimensional (1D) configuration. Most of those models have higher explosion energies than the typical value $\sim1.0\times10^{51}$\,erg for observed (note that the typical value is recently updated to $\sim0.6\times10^{51}$\,erg \cite{martinez_type_2022}).

Ref.~\cite{mori_shock_2022} performed SN simulations for $20M_\odot$ and 11.2 $\Msun$ progenitor models taking account of ALPs. Although the simulation studies \cite{mori_shock_2022,mori_multimessenger_2023} have shown that heavy ALPs help SN explosions, the number of the available models is still limited. Since a wide range of massive star with masses above $\sim10\Msun$ can lead to SN explosions \cite{sukhbold_core-collapse_2016,muller_simple_2016,ertl_two-parameter_2016,boccioli_explosion_2023}, \tkt{we need} to perform SN simulations with different progenitor models for preparing for a next Galactic SN \tkt{whose progenitor mass is unknown}. 

In this study, we perform SN simulations based on Ref. \cite{mori_shock_2022} for three progenitor stars. We use 11.2, 20 and 25 $\,\Msun$ progenitor models, taking into account ALPs with $\ma = 100-800 \,\MeV$ and $\gten= 4-10$. The ALP parameter space we explore is more extended and detailed than \tkt{that of} the previous study. \tkt{We perform 1D simulations, which do not require high computational cost, to compute 90 models in total and survey a wide range of the ALP mass, the coupling constant, and the progenitor models.}
In particular, we focus on the progenitor dependence of the ALP effects on SN dynamics.

This paper is organized as follows. In Section \ref{Methods}, we explain the ALP model we adopt and the setup of our simulations. In Section \ref{Results}, we show the results of the simulations. In particular, in Section \ref{ALP parameter dependence}, we focus on the dependence of SN dynamics on the ALP parameters for the $20\Msun$ progenitor, and in Section \ref{Progenitor dependence}, we discuss the ALP effects on different progenitors. In Section \ref{Conclusion}, we summarize  our results and make the conclusion.

\section{Methods}
\label{Methods}
Following \citet{mori_shock_2022}, 
we have performed 1D core-collapse SN simulations, taking account of ALP heating and cooling. 
For the ALP production and absorption rates, we employ the same formalism as used in \citet{mori_shock_2022}. 
In this section, we provide a brief summary of the treatment of the ALP effects in our simulations.

\subsection{ALP production rates}
\tkt{In this work, we consider a photophilic ALP model where the ALPs are generated through two photon interaction processes;}
the Primakoff process $(\gamma + p \to a + p)$ and the photon coalescence $(\gamma + \gamma \to a)$. 

The Primakoff rate is given as 
\begin{equation}\label{eq:Primrate}
    \frac{d^{2} n_{a}}{dtd\omega} \Bigg|_{\rm{prim}}  = \frac{1}{\pi^2}\omega\sqrt{\omega^{2}-\omega_{\rm{pl}}^{2}}\Gamma _{\gamma \to a} f(\omega),
\end{equation} 
where $n_a$ is the number density of ALPs, $\omega$ is photon energy and $f(\omega)$ is the Bose-Einstein distribution of photons, $\omega_{\rm{pl}}$ is plasma frequency. \tkt{The ALP energy $E$ is equal to $\omega$ because of the energy conservation, and} $\Gamma _{\gamma \to a}$ is given by \cite{di_lella_search_2000}
\begin{equation}
\begin{split}
    \Gamma _{\gamma \to a} = g^2_{a \gamma} \frac{T \kappa^2}{32 \pi} \frac{p}{E} \left(\frac{((k+p)^2+\kappa^2) ((k-p)^2+\kappa^2)}{4kp\kappa^2}\right. \times \\
    \left.\ln\left(\frac{(k+p)^2+\kappa^2}{(k-p)^2+\kappa^2}\right)- \frac{(k^2-p^2)^2}{4kp\kappa^2} \ln\left(\frac{(k+p)^2}{(k-p)^2}\right)-1\right),
\end{split}
\end{equation}
where $T$ is the temperature, $\kappa$ is the Debye-H\"{u}ckel scale, $p$ is the ALP momentum, \tkt{and} $k$ is the wave number \tkt{of} photons in plasma.

The photon coalescence rate is given as \cite{di_lella_search_2000}
\begin{equation}\label{eq:Coalrate}
    \frac{d^{2} n_{a}}{dtdE} \Bigg|_{\rm{coal}} = g^{2}_{\alpha \gamma} \frac{m^{4}_{a}}{128 {\pi}^3} p \left(1- \frac{4 {\omega}_{\rm pl}^{2}}{m^2_a}\right)^\frac{3}{2} e^{- \frac{E}{T}}.
\end{equation} 

The energy loss rates via these two processes $Q_{\cool}$ are given by 
\begin{equation}
    Q_{\rm{cool}} =\int^{\infty}_{m_{a}} d\omega \,\omega \frac{d^{2} n_{a}}{dtd\omega} \Bigg|_{\rm{prim}} +  \int^{\infty}_{m_{a}} dE\,E \frac{d^{2} n_{a}}{dtdE} \Bigg|_{\rm{coal}} .
\end{equation}
\tkt{The photon coalescence} 
 \tkt{contributes} to the ALP production only when ALPs are heavier than $2\omega_{\rm{pl}}$ where the plasma frequency $\omega_{\rm{pl}}$ is the ``effective photon mass".

\subsection{ALP absorption rate}
The ALPs produced by these processes propagate through the SN matter \tkt{and} affect the energy transfer. \tkt{If} they decay into photons \tkt{within the SN matter} and the photons are absorbed by the matter in the post-shock region, they could assist the shock revival. We consider the inverse Primakoff process $(a \to \gamma)$ and radiative decay $(a \to \gamma \gamma)$ as the ALP \tkt{decay processes.}

The inverse Primakoff rate is given as \cite{lucente_heavy_2020}
\begin{equation}
    \Gamma_{a \to \gamma} = 2 \Gamma_{\gamma \to a} / \beta_{E},
\end{equation}
where $\beta_{E}$ = $v_{a}/c$ and $v_{a}$ is the velocity of ALPs.
The radiative decay rate is given as 
\begin{equation}
    \Gamma_{a \to \gamma\gamma} = g^2_{a \gamma} \frac{m^3_a}{64\pi}\left(1-\frac{4\omega_{pl}^2}{m^2_a}\right)^{\frac{3}{2}}.
\end{equation}

\subsection{ALP transport}
The previous study \cite{mori_shock_2022} incorporated the ALP transport into SN simulations as follows.
The evolution of the ALP energy per unit volume $\mathcal{E}$ is described by
\begin{equation}\label{eq:alpevol}
    \frac{\partial \mathcal{E}}{\partial t} + \nabla \cdot \mathcal{F} = Q_{\rm{cool}} - Q_{\rm{heat}} ,
\end{equation}
where $\mathcal{F}$ is the ALP energy flux and $Q_\textrm{heat}$ is the heating rate per unit volume due to ALPs. Assuming stationarity and spherical symmetry, Eq.~(\ref{eq:alpevol}) is simplified to
\begin{equation}
    \frac{1}{4 \pi r^{2}} \frac{\partial}{\partial r}(L) = Q_{\rm{cool}} - Q_{\rm{heat}},\label{eq:alpevol2}
\end{equation}
here $L$ is ALP luminosity. We solve Eq.~(\ref{eq:alpevol2}) at each time step to obtain $Q_\mathrm{heat}$. We then update the internal energy at the $n$-th step as
\begin{eqnarray}
    e_\mathrm{int}^{n+1}=e_\mathrm{int}^n+(Q_\mathrm{heat}-Q_\mathrm{cool})\Delta t,
\end{eqnarray}
where $e_\mathrm{int}$ is the internal energy and $\Delta t$ is the time step. 

\subsection{Model setup}
The numerical setup in this study is essentially the same as in \cite{mori_shock_2022}.
We employ the 3DnSNe code \cite{takiwaki_three-dimensional_2016}, which is a multi-dimensional neutrino radiation hydrodynamics code developed to study core-collapse SNe. 
The neutrino transport is solved by the three-flavor isotropic diffusion source approximation (IDSA) scheme \cite{liebendoerfer_isotropic_2009}. 
We use the state-of-the-art neutrino opacity \cite{kotake_impact_2018} and the neutrino energy spectrum is discretized with 20 energy bins for $0 < \varepsilon_\nu \leq 300 \,\MeV$. 
We take account of the effective general relativistic effect \cite{marek_exploring_2006} for the gravitational potential and the  gravitational redshift for the neutrino transport.

\begin{figure}[t]
    \centering
    \includegraphics[width=8cm,clip]{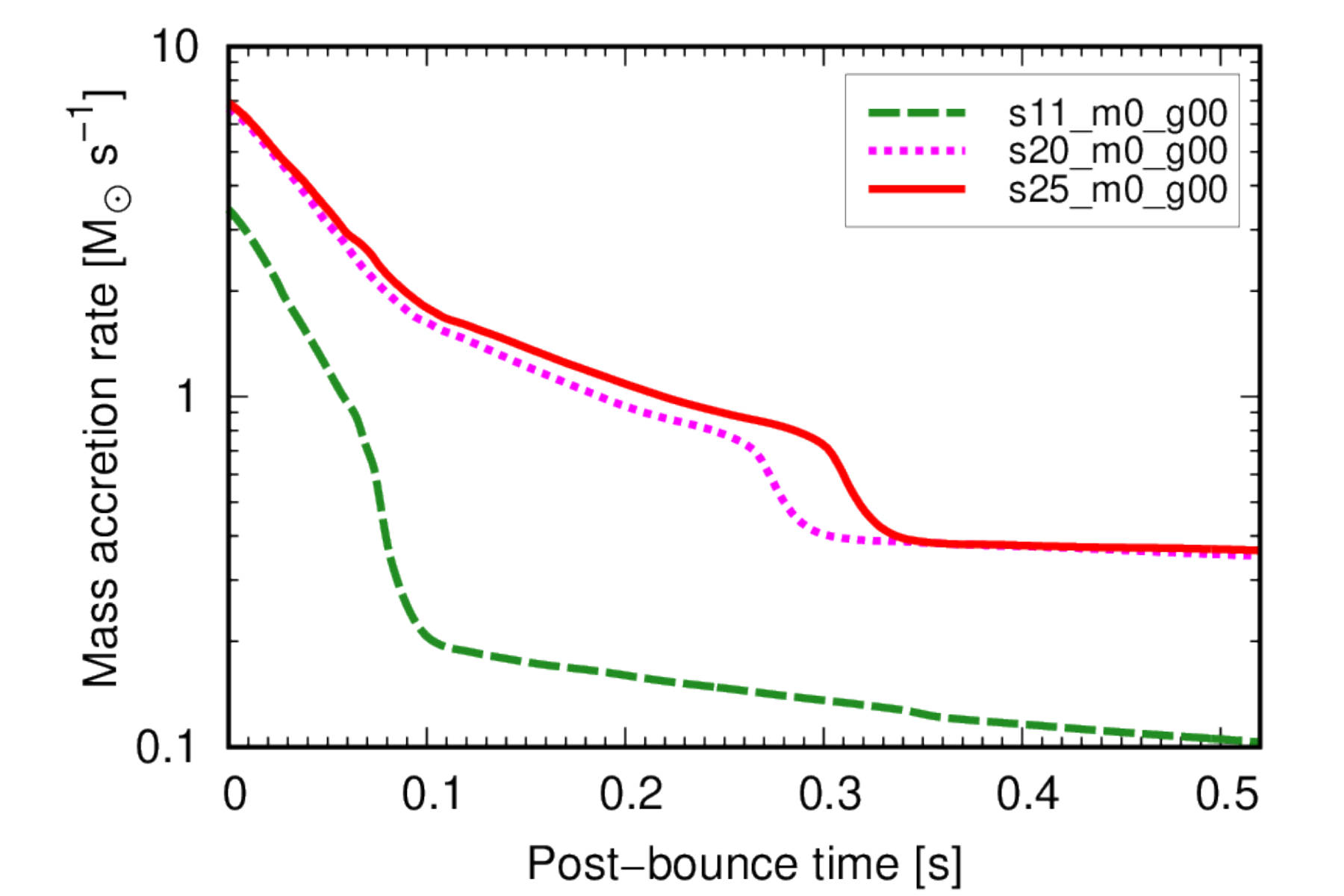}
    \caption{The mass accretion rate, measured at $r=500\,\rm{km}$, is shown as a function of time after the core bounce $\tpb$ for the no-ALP models. Here, the solid red, dotted magenta, and dashed green lines represent the $11.2\Msun$, $20\Msun$ and $25\Msun$ progenitor models. The accretion rate decreases over time in all three models, with a pronounced drop when the Si/O interface falls onto the central region. 
}
\label{fig:macc}
\end{figure} 

We use the 11.2$\,\Msun$ progenitor model from \cite{woosley_evolution_2002}, and the 20$\,\Msun$ and 25$\,\Msun$ models from \cite{woosley_nucleosynthesis_2007}, 
to investigate the progenitor dependence of ALP effects on core-collapse SN explosions.
There are two reasons for the choice of these progenitors.
First, these three progenitor models are red supergiants and massive enough 
\tkt{for their core to undergo gravitational collapse during the final stage of their evolution} 
\cite{smartt_death_2009,smartt_detection_2004}. 
Their explosions will appear as type I\hspace{-.1em}I SNe, 
\tkt{the most commonly observed type among core-collapse SNe}. 
Second, these progenitor models have a different mass-accretion history, \tkt{due to their differences in compactness}, as shown in Fig.\,\ref{fig:macc}. 
Since the gravitational potential of the accreting gas is the dominant energy reservoir for neutrino-driven explosions, the difference in the mass accretion rate leads to different dynamical evolution. 
Moreover, the thermal evolution of their core (for example, the maximum temperature achieved in the core) strongly affects the ALP production rate. 
The models examined in this study are suitable for the purpose to inspect the dependence of the ALP emissions on the progenitor structure.

For these progenitor models, we performed one-dimensional core-collapse SN simulations to investigate the impact of ALPs on the core-collapse SN dynamics. The ALP parameter space has been constrained by various studies \cite[e.g.][]{
lucente_heavy_2020, lucente_constraining_2022, PhysRevD.98.055032, 
na64_collaboration_search_2020, carenza_constraints_2020, D_brich_2020, dolan_revised_2017, Diamond_2024, dev2024constraintsphotoncouplingaxionlike}.
ALPs with $m_a=\mathcal{O}(100)$\,MeV are reported to have significant effects on core-collapse SN dynamics \cite{Caputo_muonic_2022, mori_shock_2022}.
Considering these results, we have chosen the parameter space of ALP mass, $\ma = 100-800\,\MeV$, and ALP-photon coupling constant, $\gten = g_{a \gamma} / (10^{-10} \,\rm{GeV^{-1}}) = 4-10$. 
Note that there is also discussion that excludes some parts of this range \cite{Caputo_low-energy_2022, Fiorillo:2025yzf, Diamond_2024}.

The models are labeled as follows. For example, a model with the $20.0\,\Msun$ progenitor and ALPs with $\ma=200\,\MeV$ and $\gten=8$ is called `s20\_m2\_g08'.  In addition, the model without ALPs is labeled as `s20\_m0\_g00' and referred to as the no-ALP model.

\section{Results}
\label{Results}
\tkt{We perform 90 simulations for the three progenitor models until the post-bounce time $\tpb=0.52$\,s.
In Section \ref{ALP parameter dependence}, we present the properties of the $20\Msun$ models and examine their dependence on the ALP parameters, $m_a$ and $g_{a\gamma}$.
In Section \ref{Progenitor dependence}, we compare the characteristics of the models with different masses for some certain choices of the ALP parameters.}

\subsection{ALP parameter dependence}
\label{ALP parameter dependence}

When ALPs are not taken into account, shock revival does not occur in one-dimensional core-collapse SN models \cite[e.g.][]{liebendorfer_probing_2001,2018JPhG...45j4001O}, except for models with a low-mass progenitor \cite{2006A&A...450..345K,fischer_protoneutron_2010,hudepohl_neutrino_2010}.
However, in some of our models considering ALPs, the shock wave is successfully revived and leads to a supernova explosion.

\begin{figure}[t]
    \centering
    \includegraphics[width=8cm,clip]{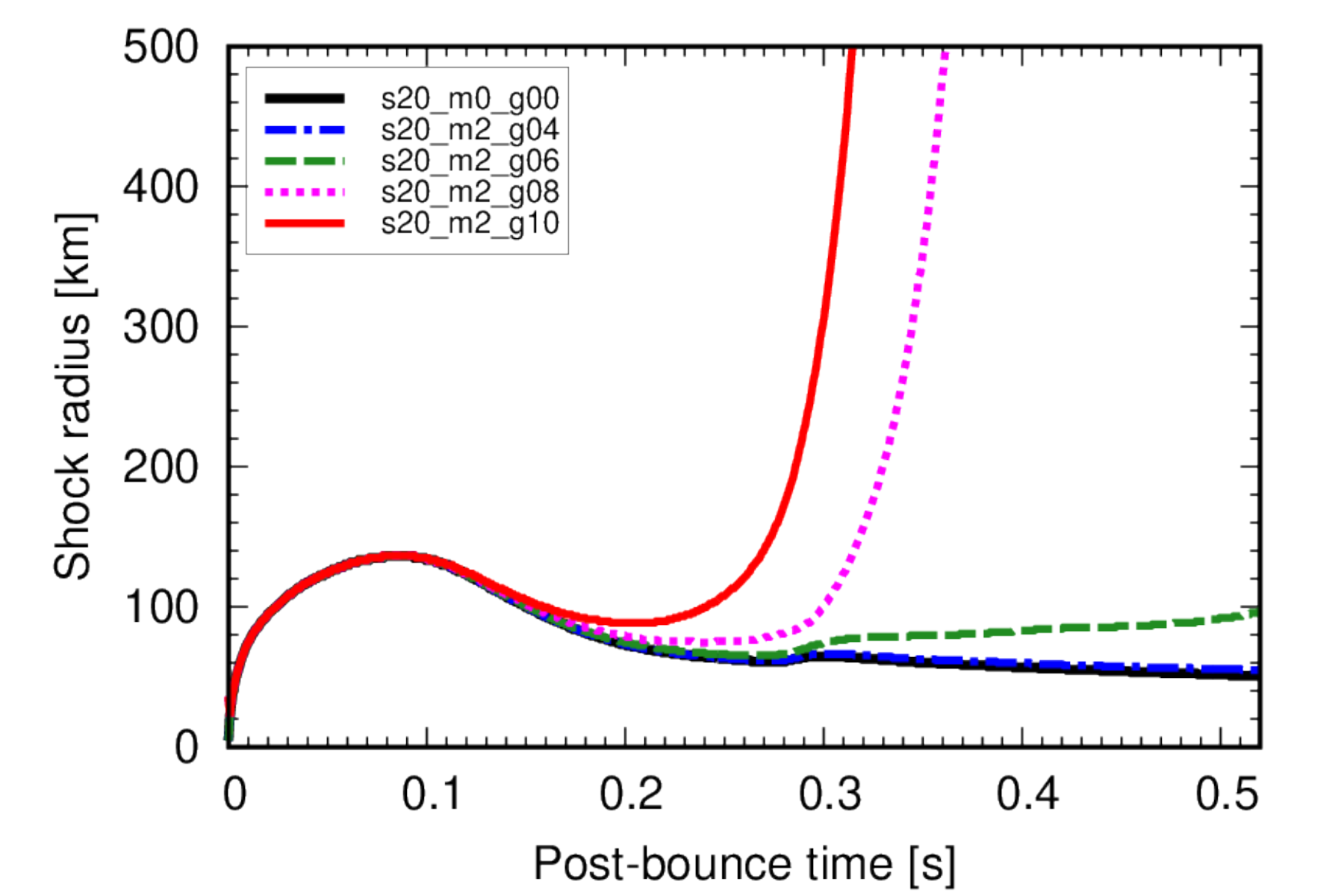}
    \caption{\label{fig:sh20_200} The angular-averaged shock radii as functions of time for the s20\_m2  model series. Shown are the models with ALPs with different coupling constants $\gten=4,6,8\,\rmand\,10$, as labeled, as well as the no-ALP model (black solid line) represented as s20\_m0\_g00. For the s20\_m2\_g08 and s20\_m2\_g10 models (solid red and dotted magenta lines), the additional heating via ALP-photon interactions leads to the shock revival. Shock wave is more likely to be revived in models with higher coupling constants.} 
\end{figure}

Figure \ref{fig:sh20_200} shows the time evolution of the shock radius for the s20\_m2 model series with different coupling constants $g_{10}=4$, 6, 8 and 10, along with the no-ALP model denoted by s20\_m0\_g00. 
For the no-ALP \tkt{model and the $g_{10}=4$ model (solid black and dash-dot blue lines),} the shock wave stalls at $r\sim 150\,\km$ and fails to revive. This behavior is similar to conventional one-dimensional simulations without ALPs.
However, 
\tkt{in the cases of the $g_{10}=8$ and 10 models (dotted magenta and solid red lines), the shock wave successfully revives and turns to a runaway expansion around $\tpb=0.3$\,s.}
For the \tkt{$g_{10}=6$ model} (dashed green line),
the shock wave \tkt{gradually expands after $t_{\rm pb} \sim 0.3$ s, but the shock radius remains at 100 km at the end of the simulation and it is not clear if the shock expansion continues afterward.}
These models indicate \tkt{the trend} that the shock wave is more likely to revive \tkt{for the models} with higher coupling constants. 

\begin{figure}[t]
    \centering
    \includegraphics[width=8cm,clip]{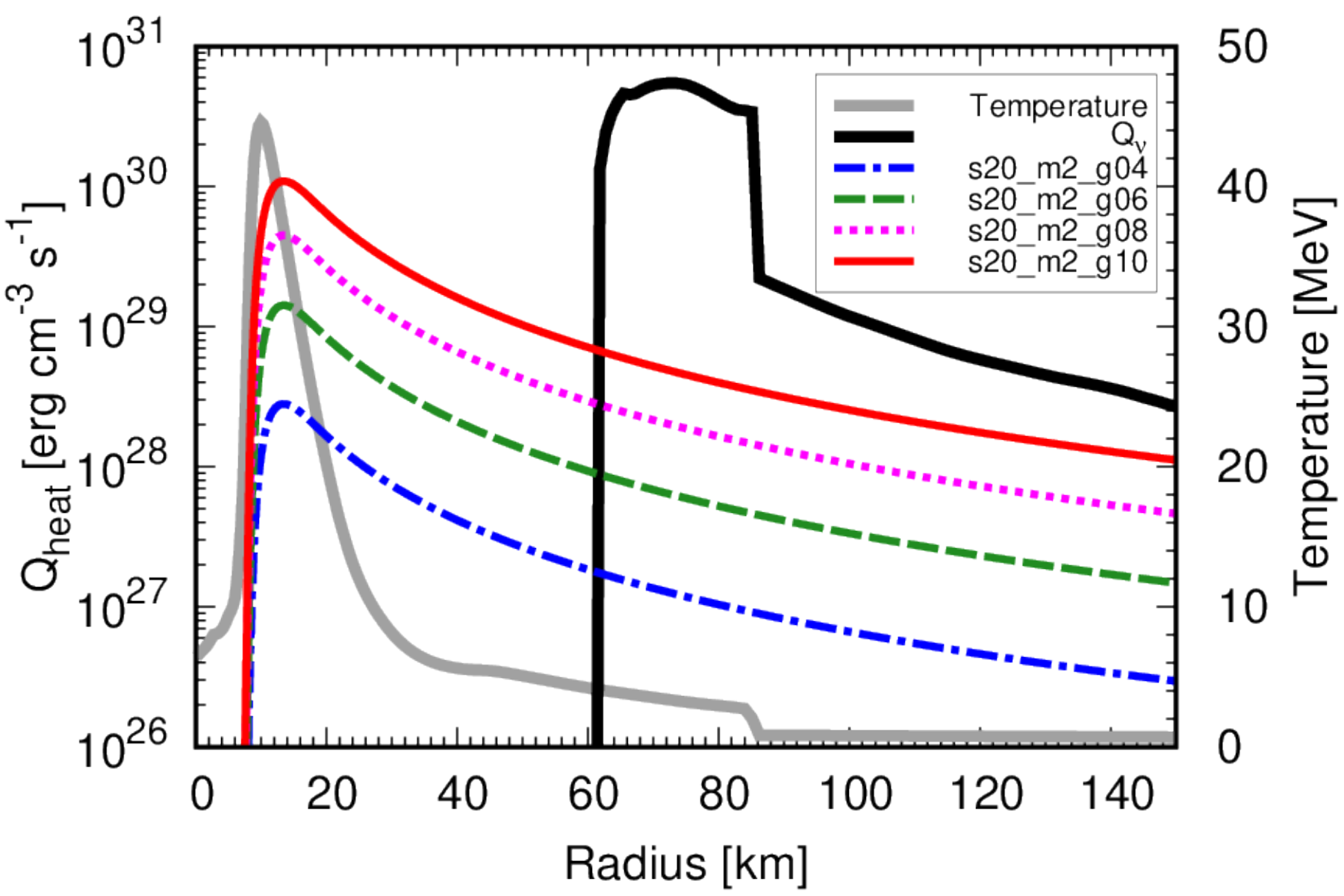}
    \caption{\label{fig:ht20_200} The radial profile of the ALP heating rate $Q_{\heat}$ at $\tpb= 170 \, \rm{ms}$ is shown.
    The color coding is the same as in Fig.\ref{fig:sh20_200}. That said, the black solid line represents the net neutrino heating rate $Q_{\nu} = Q_{\heat}^{\nu} - Q_{\cool}^{\nu}$, and the gray solid line represents the SN matter temperature for the s20 no-ALP model. A higher coupling constant leads to higher ALP heating rates, facilitating shock revival. For these ALP parameter sets, the neutrino heating rates dominate over the ALP heating rates in the gain region ($r\sim 60-80$ km), and neutrino heating is the main explosion mechanism.}
\end{figure}

\tkt{This trend} 
can be explained by the dependence of the ALP heating rate $Q_\mathrm{\heat}$ induced by ALP decay photons on the coupling constant. In Fig. \ref{fig:ht20_200}, the radial profiles of $Q_{\heat}$ are shown for the s20\_m2 model series at the time $\tpb=0.17$\,s.
We can see that a higher coupling constant induces higher ALP heating rates, which can lead to the successful shock revival.

Figure \ref{fig:ht20_200} shows that the ALP heating rates peak at $r \sim10$ km regardless of the coupling constant. This is because the temperature distribution (drawn in the solid gray line) peaks at this location, where the ALP production rate is also maximized due to its sensitivity to temperature.
Furthermore, because of the ray-by-ray approximation used in solving ALP transport, ALPs cannot propagate inward from the radius where they are produced. As a result, there is a precipitous cutoff in $Q_{\heat}$ \tkt{at} $r\sim10$ km. 
\tkt{Outside of the peak, $Q_{\heat}$ decreases following a $1/r^2$ dependence shown in Eq. (\ref{eq:qheat}).}

\tkt{We can also see that the neutrino heating rate (drawn in the solid black line) is dominant between the gain radius $R_{\rm{gain}} \sim 60$ km and the the shock radius $R_{\rm{shock}} \sim 85$ km. 
The neutrino cooling dominates over the heating inside the gain radius, and they balance out at $r=R_{\rm{gain}}$.
The neutrino heating rate drops at} \tkt{$r=R_{\rm{shock}}$} \tkt{because nucleons, which are the primary targets of the neutrino absorption process, are scarcely present outside the shock radius.}
\tkt{The excess of the neutrino heating rate over the heating rate of ALPs implies that supernova explosions are primarily driven by the neutrino heating and} 
ALPs play a \tkt{secondary} role.

\begin{figure*}[t]
    \centering
    \includegraphics[keepaspectratio, width=17.5cm,clip]{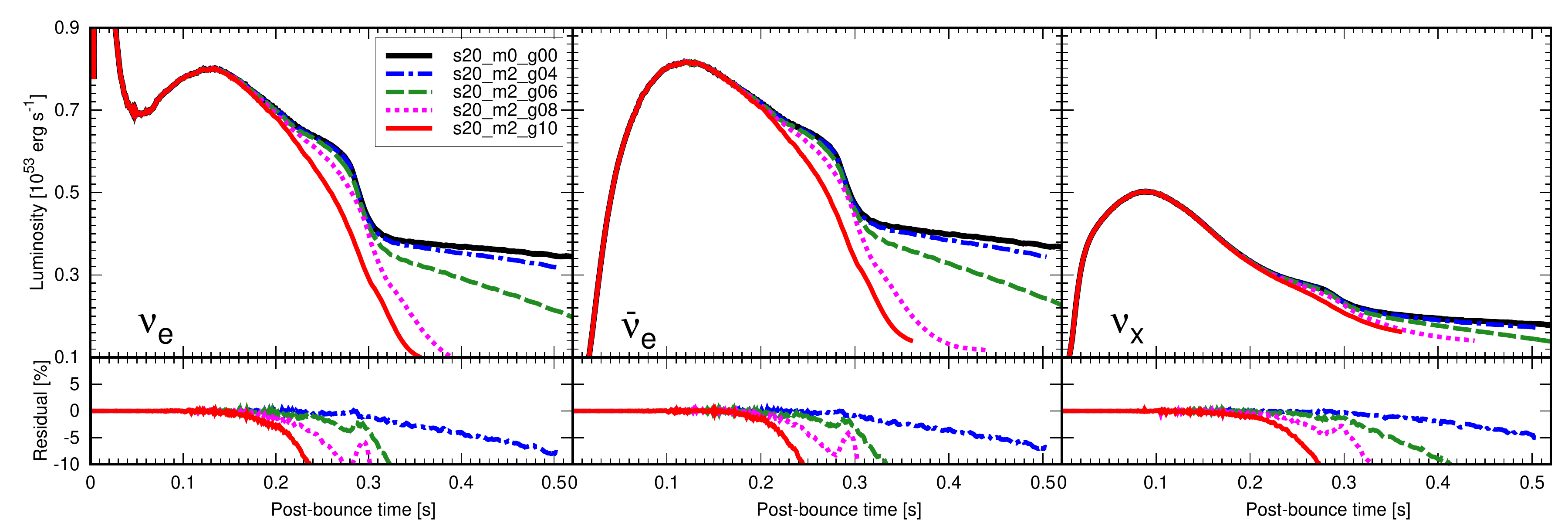}
    \includegraphics[keepaspectratio, width=17.5cm,clip]{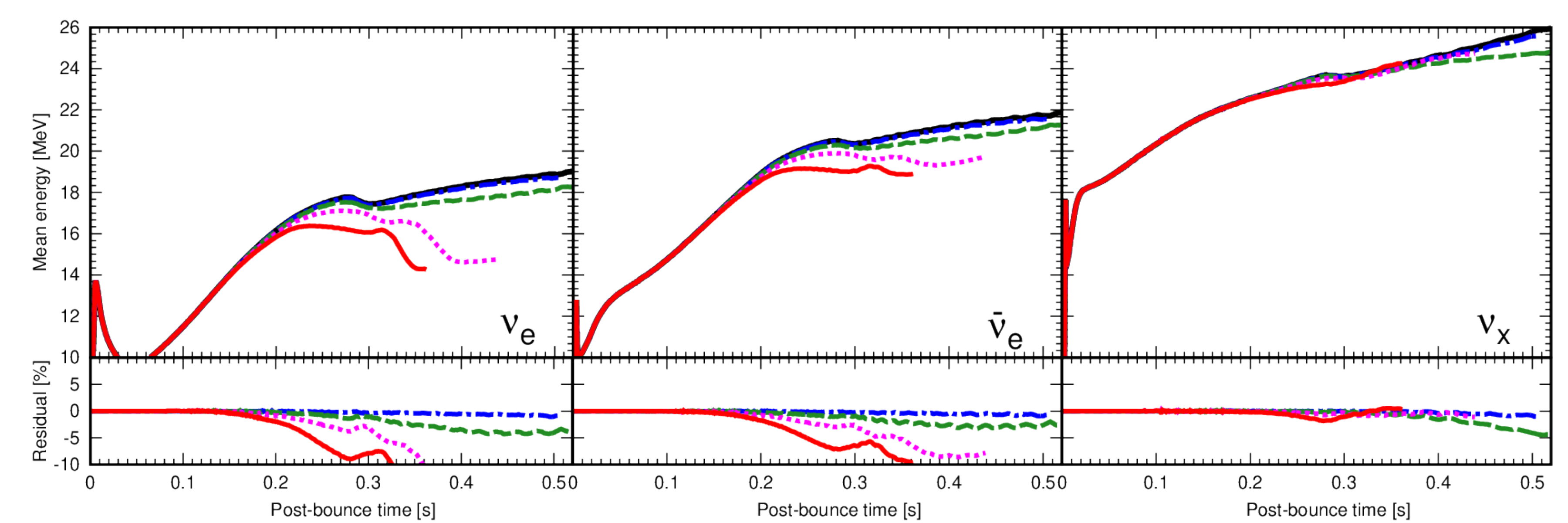}
  \caption{\label{fig:LumEne} 
    The neutrino luminosity (top panels) and the mean energy (bottom panels).
    The residuals relative to the no-ALP models are shown below each plot. The color coding is the same as in Fig.\ref{fig:sh20_200}.
    The higher coupling constant model shows that a greater reduction in the luminosity of all flavors and in the mean energy of $\nu_{e}$ and $\bar{\nu}_{e}$, because the shock expansion leads to a decrease in the mass accretion rate. 
    On the other hand, the mean energy of $\nu_{x}$ is almost independent of the models. This is because $\nu_{x}$ is produced more in the central region than the neutrinos of other flavors, and $\nu_{x}$ is less influenced by the shock expansion. Before the shock expansion, the luminosity and the mean energy are not affected by ALPs, but they are influenced by the shock expansion, which is promoted by ALPs.
  }
\end{figure*}

In Fig. \ref{fig:LumEne}, we present the neutrino luminosity and mean energy for the s20\_m2 model series, along with the residuals relative to the no-ALP model.
Before shock revival \tkt{($\tpb= 0.2-0.3$\,s)}, 
\tkt{the luminosity and the mean energy for all flavors are essentially unchanged by considering the effects of ALPs.}
However, \tkt{the models with higher coupling constants show decrease in the $\nu_{e}$ and $\bar{\nu}_{e}$ luminosities and their mean energies, because these models experience early shock revival,} leading to a decrease in the mass accretion rate.
\tkt{Only a slight} difference is observed in the $\nu_{x}$ mean energy among these models because $\nu_{x}$ is \tkt{predominantly} produced 
\tkt{in the dense core} via neutral current pair production \cite[e.g.][]{betranhandy_impact_2020} 
\tkt{and less affected by the mass accretion history.}
Although ALP emission can rapidly extract energy from the PNS, it takes \tkt{a time} on the order of the neutrino diffusion time (several seconds) for the information to reach the neutrino sphere \cite{PhysRevD.101.123025}.

\begin{figure}[t]
    \includegraphics[width=8cm,clip]{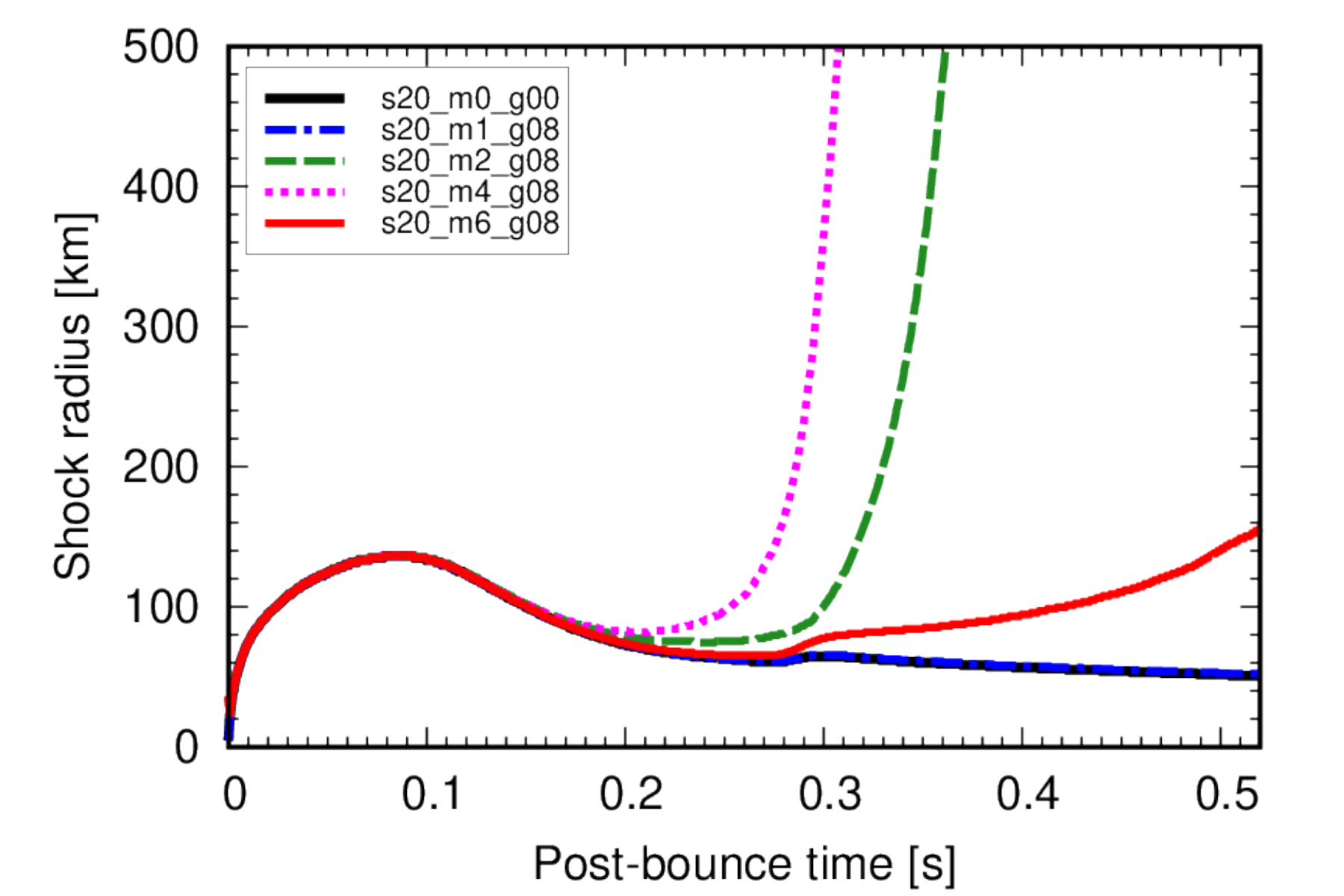}
    \caption{\label{fig:sh20_08} The shock radius of the s20\_g08  model series. Shown are the models with ALPs with different masses $(\ma=100,\, 200,\, 400\, \rm{and}\,600 \,\MeV)$, as labeled, along with the model without ALPs (black solid line). The s20\_m2\_g08 and s20\_m4\_g08 models (dashed green and dotted magenta lines) exhibit shock revival. However, for the s20\_m6\_g08 model (solid red line) in which heavier ALPs are incorporated, the shock expansion is delayed compared to these two models. The tend of shock revival is not monotonic with respect to the ALP mass.
    }
\end{figure}

The SN dynamics is also dependent on the ALP mass.
In Fig. \ref{fig:sh20_08}, we plot the shock evolution for the s20\_g08 model series with $\ma=100$, 200, 400, and 600 MeV. \tkt{For the model with $m_a = 100$ MeV (dash-dot blue line),} 
the shock wave shows almost no change compared to the no-ALP model and fails to revive, whereas 
\tkt{the shock waves of the models with $m_a =$ 200 and 400 MeV models (dashed green and dotted magenta lines) turn to expand at $\sim 300$ ms and 250 ms after bounce.} 
\tkt{Therefore, within} the range of $\ma=100-400 \,\MeV$, higher ALP masses lead to an earlier 
\tkt{shock revival.}
However, in the case of 
\tkt{the model with $m_a = 600$ MeV (solid red line),} the propagation of the bounce shock is slower than in the models with lighter ALPs, and in the \tkt{$m_a = 800$ MeV} model, the shock wave does not initiate \tkt{a runaway expansion,} similar to the no-ALP model (solid black line).
These results indicate that a higher ALP mass does not necessarily facilitate the shock revival and that the relationship between the ALP mass and the shock wave \tkt{behavior} exhibits a non-monotonic trend.

\begin{figure}[t]
    \centering
    \includegraphics[keepaspectratio, width=8cm,clip]{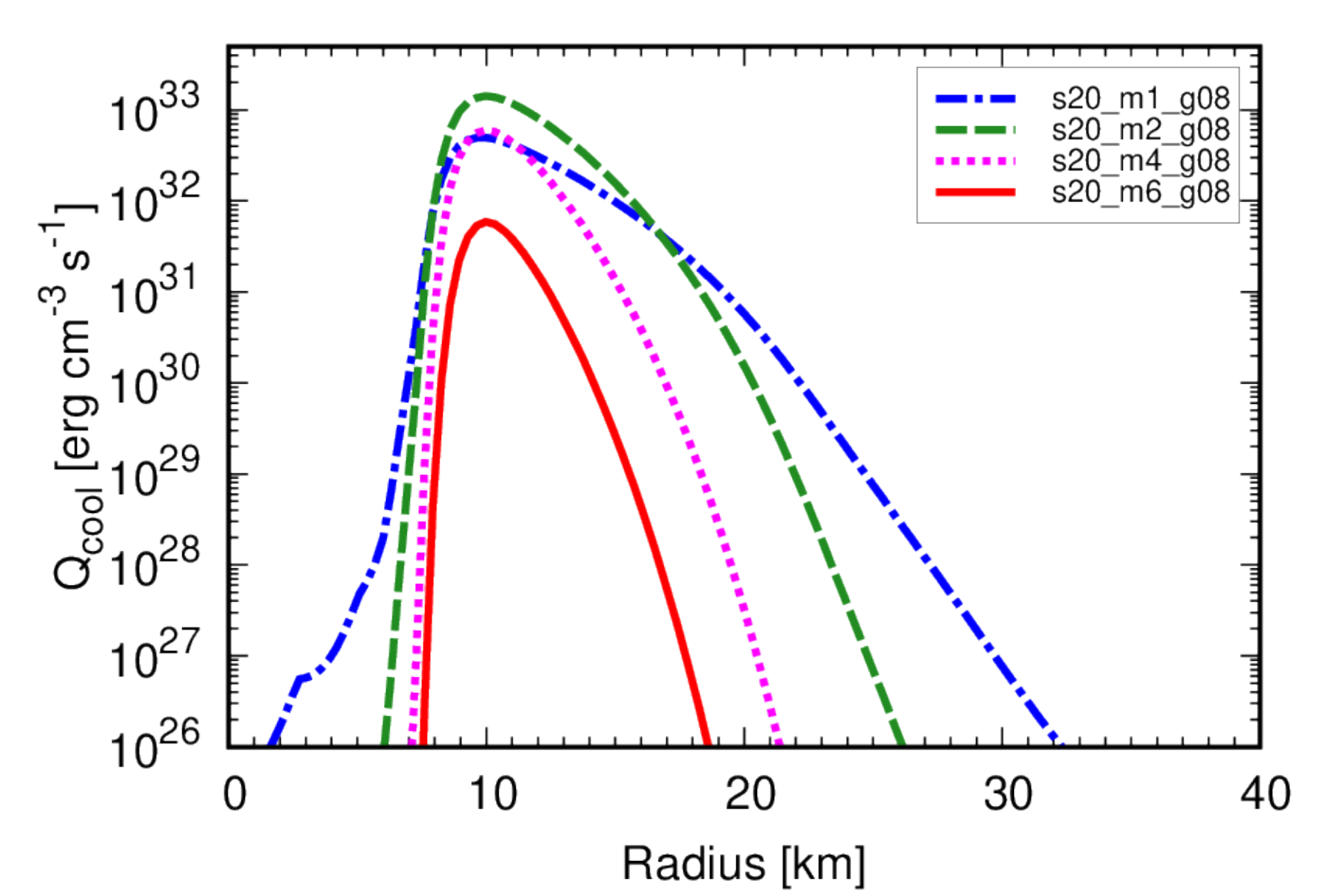}
    \caption{\label{fig:cl20_08} The radial profile of the ALP cooling rate at $\tpb = 170\,\rm{ms}$.
    The color coding is the same as in Fig.\ref{fig:sh20_08}. 
    The ALP production rate peaks at $r\sim10$ km for all models, since the SN matter temperature also peaks at the same location. In the case of the s20\_m6\_g08 model, the peak is about one order of magnitude lower than in the other models. This is due to the temperature not being sufficient for 600 MeV ALPs to be produced. }
\end{figure}

The non-monotonic dependence on $m_a$ can be explained by the \tkt{$m_a$ dependence of the} ALP production rate. Fig. \ref{fig:cl20_08} shows the radial profile of $Q_\mathrm{cool}$ at $\tpb=0.17$\,s. 
The cooling rates reach a peak at $r\sim 10 \,\km$ regardless of the ALP masses, because the ALP production processes are sensitive to the matter temperature, which also peaks in this region. 
In the case of the s20\_m6\_g08 model (and the s20\_m8\_g08 model as well), the peak \tkt{value} is about an order of magnitude lower than \tkt{the value} in the other models. 
This is due to the Boltzmann suppression, as the temperature is not high enough to produce 
\tkt{such heavy ALPs.} 

 \begin{figure}[t]
    \includegraphics[keepaspectratio, width=8cm,clip]{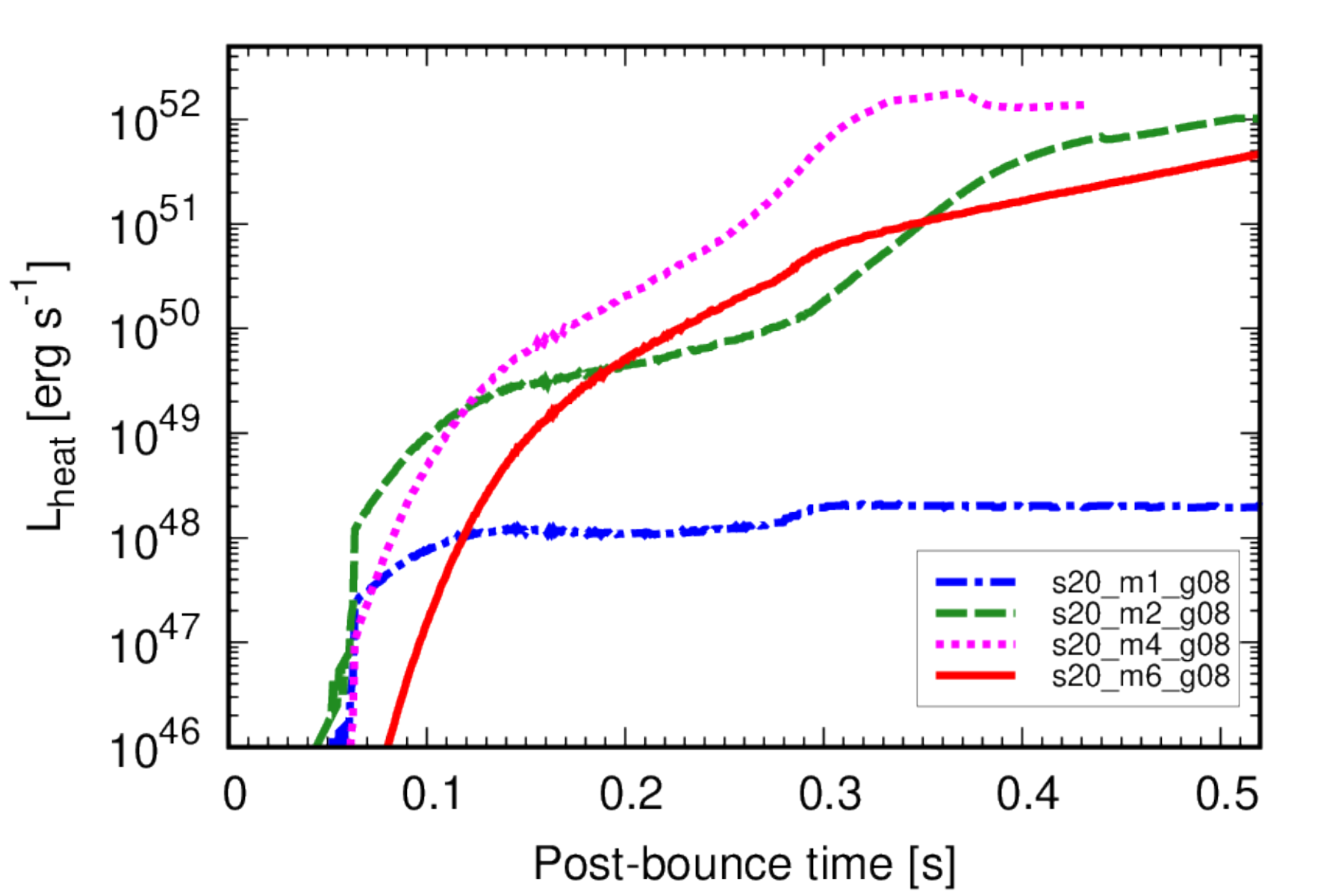}
    \caption{\label{fig:ht20_08} The ALP heating rate $L_\mathrm{heat}$ in the gain region.
    The color coding is the same as in Fig.\ref{fig:sh20_08}. For the s20\_m2\_g08 and s20\_m4\_g08 models (dashed green and dotted magenta lines), the heating rate increases significantly from $\tpb \sim 0.05$\,s onward. However, for the s20\_m6\_g08 model (solid red line), the heating rate increases with a delay at $\tpb\sim0.08$\,s compared to the other models, as it takes longer to reach the temperature sufficient for 600 MeV ALPs to be produced. Therefore, the s20\_m6\_g08 model exhibits a delay in shock expansion.
    }
\end{figure}

\begin{figure*}[t]
    \centering
    \includegraphics[keepaspectratio, width=14cm,clip]{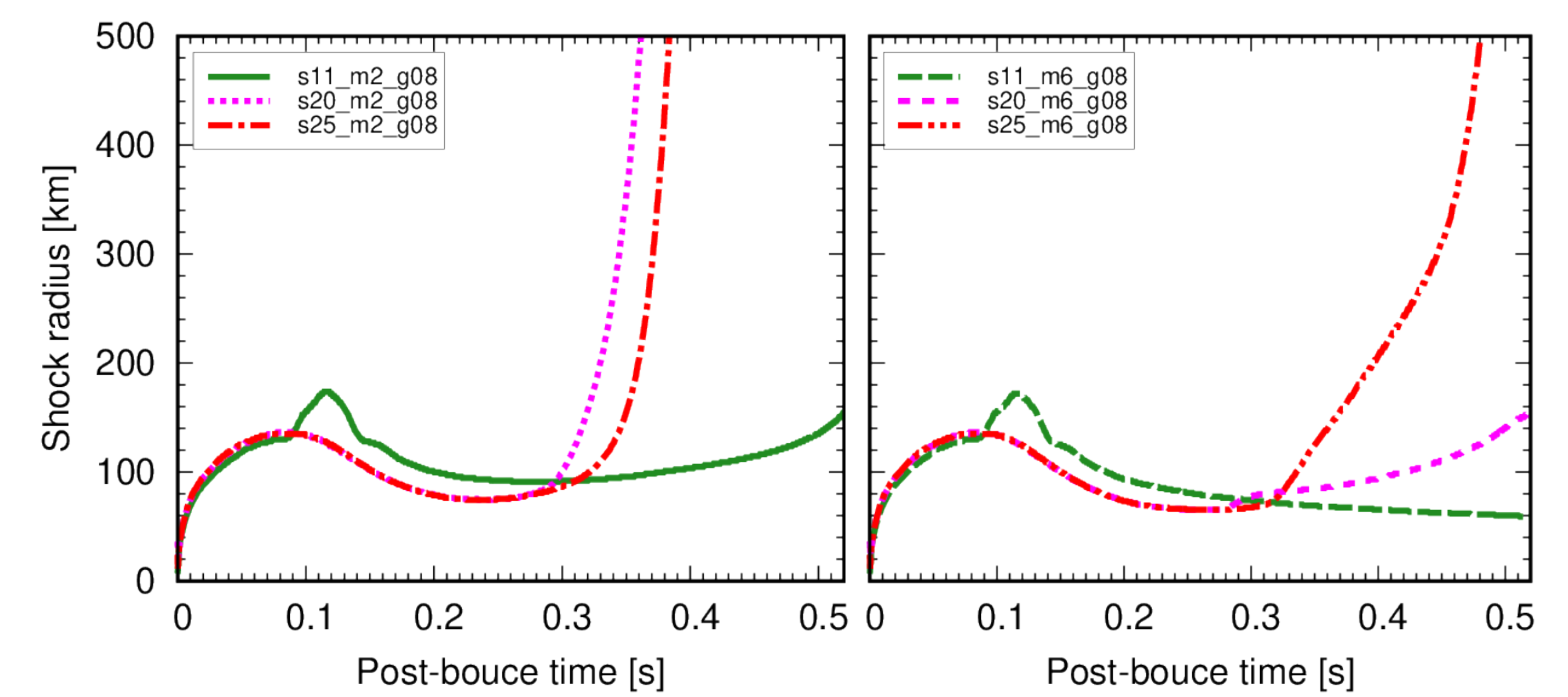}
  \caption{\label{fig:Sh_progen} The time evolution of the shock radius in the s11\_g08, s20\_08 and s25\_g08 model series with different ALP masses ($\ma= 200\,\rm{and}\, 600 \,\MeV$). 
  In the s11\_m2\_g08 model (solid green line), the shock wave gradually expands, and the s20\_m2\_g08 and s25\_m2\_g08 models (dotted magenta and dash-dot red lines) show shock revival (left panel). 
  While the s25\_m6\_g08 model (dash-dot-dot red line) shows shock revival, the shock wave in the s20\_m6\_g08 model (loosely dashed magenta line) transitions to expansion without reviving (right panel). Moreover,for the s11\_m6\_g08 model (dashed magenta line), the shock wave stalls. Among these models, shock revival is more likely in heavier progenitor stars, with this trend being particularly prominent for heavier ALPs, such as 600 MeV.  }
\end{figure*}

Figure \ref{fig:ht20_08} shows the time evolution of the ALP heating rate in the gain region defined as
\begin{eqnarray}
L_{\heat}=4\pi\int^{r_{\rm{sh}}}_{r_{\rm{gain}}} r^2Q_{\heat} \,dr.
\end{eqnarray}
For the s20\_m1\_g08 model (dash-dot blue line), the ALP heating rate is as low as $L_{\heat}\sim 10^{48}$\,erg s$^{-1}$, which is insufficient for the revival of the shock wave.
The s20\_m2\_g08 and s20\_m4\_g08 models (dashed green dotted magenta lines) show higher heating rates of $L_{\heat}\sim 10^{50}$\,erg s$^{-1}$ at $\tpb= 0.1$\,s compared to the other models, thus photons from decaying ALPs effectively heating the region behind the shock, promoting shock revival.
The differences in the heating rates among these models can be explained by the fact that heavier ALP masses contribute to the higher heating rate and more efficient heating in the gain region because of their shorter mean free path.
However, in the s20\_m6\_g08 model (solid red line), despite the higher mass, the heating rate $L_{\heat}$ is lower than that of the s20\_m2\_g08 and s20\_m4\_g08 models due to a reduction in the production rate.
Furthermore, the s20\_m6\_g08 model shows a delayed increase in the heating rate compared to the other models, because it takes longer to reach a temperature 
\tkt{to sufficiently}
produce 600 MeV ALPs. These delays in the increases in heating rate and temperature lead to the delayed shock expansion.

\subsection{Progenitor dependence}
\label{Progenitor dependence}

In this section, we discuss the progenitor dependence of the shock evolution and its underlying causes, focusing on the s11\_g08, s20\_g08 and s25\_g08 model series with ALP masses $\ma= 200$ and $600\,\MeV$.

Figure \ref{fig:Sh_progen} shows the shock evolution for each model. 
For the $11.2M_\odot$ progenitor model, a bump in the shock radius is observed at $\tpb \sim 0.1$\,s, which is induced by the infall of the oxygen-silicon layer.
\tkt{For the models with $m_a=200$ MeV (the left panel of Figure \ref{fig:Sh_progen}), the $11.2M_\odot$ model (solid green line)} 
shows slow outward propagation of the shock wave, although it is not clear whether the model successfully explodes. 
\tkt{The $20 M_\odot$ and $25 M_\odot$ models (dotted magenta and dash-dot red lines) show a successful shock revival at $\tpb \sim 0.3$\,s.}

Figure \ref{fig:Sh_progen} also shows that 
\tkt{the shock expansion is suppressed for the models with heavy ALPs ($m_a = 600$ MeV, the right panel). This is } because \tkt{the production rate of heavy ALPs is} 
relatively low. 
In particular, the s20\_m6\_g08 model (loosely dashed line) exhibits a pronounced delay in shock expansion, with the shock failing to reach $200\,\km$ by $\tpb \sim 0.52$\,s.
These results show that 
\tkt{the heavier progenitor models facilitated by ALPs are more likely to undergo shock expansion.}
This trend becomes more pronounced for \tkt{the models with} heavier ALPs such as $600$\,MeV.

\begin{figure}[t]
    \centering
    \includegraphics[width=8cm,clip]{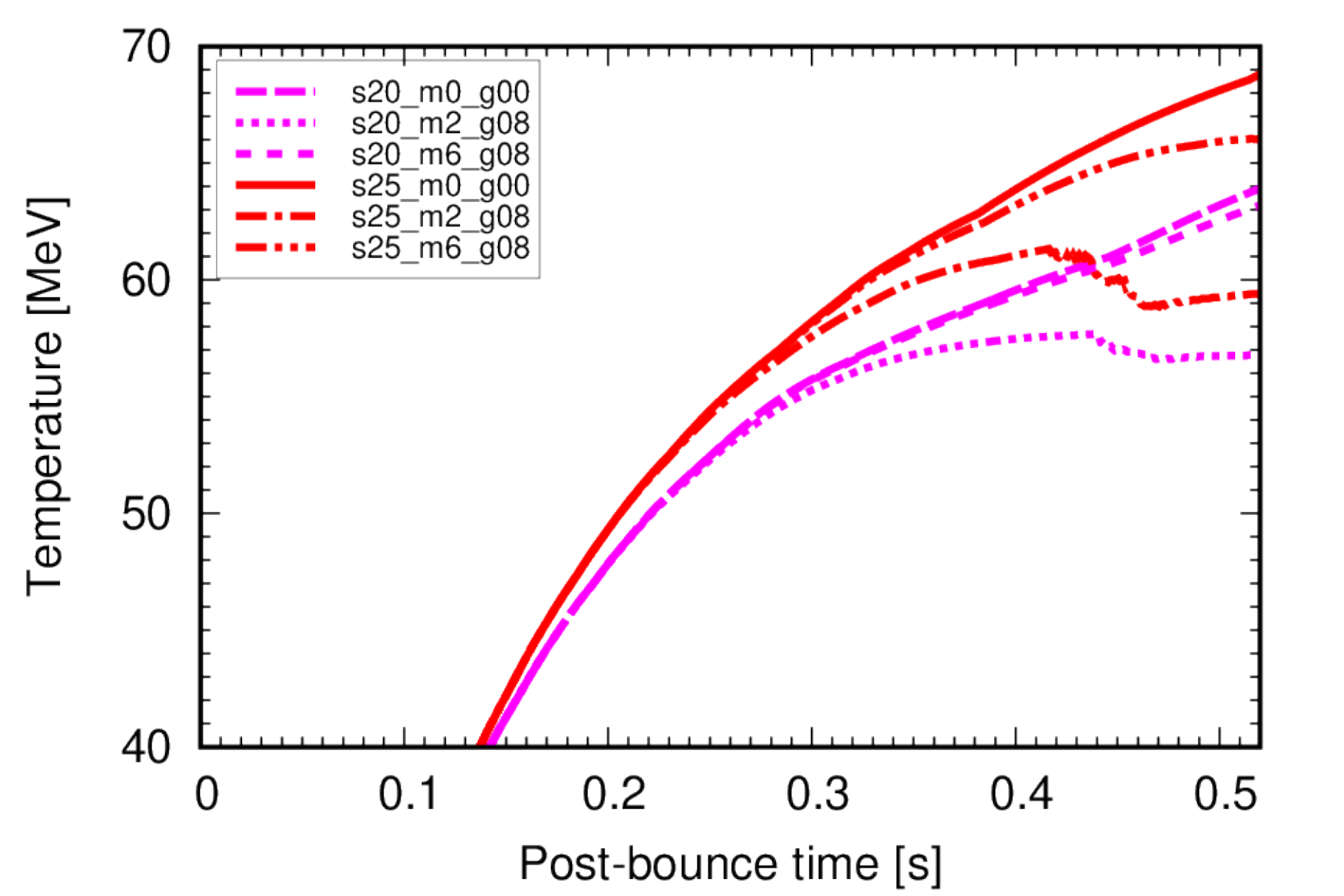}
    \caption{\label{fig:tp10} The time evolution of the maximum temperature in the PNS for the s20 and s25 series models. Here, the no-ALP models are also plotted for comparison.
    The temperature increases over time due to the gravitational compression. For the same ALP parameter set, the s25 models (red lines) show higher temperatures than the s20 models (magenta lines).
    }
\end{figure}

The dependence of the shock behavior on the progenitors is induced by the difference in the maximum temperature of the SN models.
Fig.~\ref{fig:tp10} shows the maximum temperature in the SN core, which is achieved at $r\sim10$\,km, with a focus on the s20 and s25 \tkt{model series}.
For \tkt{all} models, the temperature increases over time and reaches $55-60\,\MeV$ at $\tpb = 0.3$\,s \tkt{due to}
the gravitational compression of the accretion layer associated with the growth of the PNS mass. 
After $\tpb = 0.3$\,s, \tkt{the models taking account of the ALP effects}
show that the increase in temperature is suppressed because the shock expansion \tkt{achieved at this time} causes a decrease in the accretion rate. 

\tkt{Figure~\ref{fig:tp10}}
also shows that, for any ALP parameter sets, the peak temperature of the s25 series models (red lines) is higher than that of the s20 series models (magenta lines).
\tkt{The difference in the PNS temperature can be interpreted 
 in terms of the positive correlation between the iron core mass and the compactness parameter \cite{O'Connor_2011}. 
 For massive or high-compactness progenitors, the 
 iron core masses tend to be heavier due to the thermal pressure support. This explains a general tendency that a  progenitor with higher compactness leads to a hotter PNS. }

\tkt{The ALP production rate $Q_\mathrm{cool}$ is higher for a heavier progenitor model because of the higher temperature. As a result, the ALP heating rate $Q_\mathrm{heat}$ becomes also higher for a heavier progenitor.}

\begin{figure*}[t]
    \centering
    \includegraphics[keepaspectratio, width=14cm,clip]{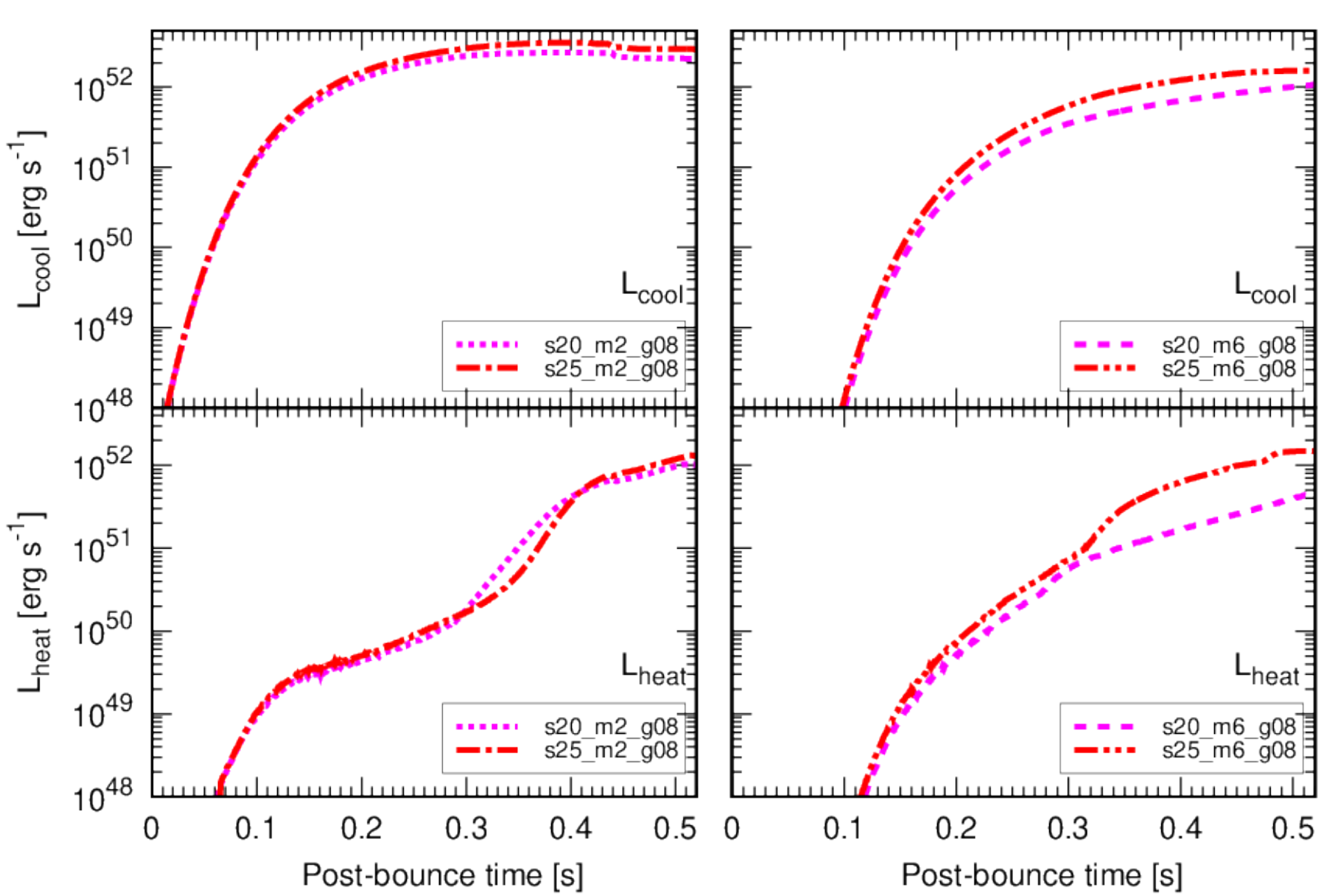}
    \caption{\label{fig:HC_progen} 
    The time evolution of the ALP heating and cooling rate, $L_\mathrm{heat}$ and $L_\mathrm{cool}$.
    The color coding is the same as in Fig.\ref{fig:tp10}.
    Before the shock revival ($\tpb \sim 0.3$\,s), the ALP cooling and heating rates show a relatively small difference between the s20\_m2\_g08 and s25\_m2\_g08 models (left panel, dotted magenta and dash-dot red lines). Moreover, for the s25\_m6\_g08 model (right panel, dash-dot-dot red line), the cooling and heating rates are obviously higher than those for the s20\_m6\_g08 (loosely dashed magenta line). This is because it is realized the s25 models reach higher temperatures than the s20 models. 
    Therefore, in heavier progenitor models tend to show more effective ALP heating rates, which contribute to shock revival. 
    }
\end{figure*}

In Fig. \ref{fig:HC_progen}, we show the evolution of the ALP cooling and heating rates in the gain region. 
Here the cooling rate is defined as 
\begin{eqnarray}
L_{\cool}=4\pi\int^{\rm{r_{sh}}}_{0} r^2Q_{\cool}dr.
\label{L_cl}
\end{eqnarray}
For the m2 model series, the cooling rates begin to increase early, with little noticeable difference between the 
\tkt{$20\,\Msun$ and $25\,\Msun$ models (dotted magenta and dash-dot red lines).} 
This is because the temperature required to produce 200\,MeV ALPs is comparatively lower than that for heavier ALPs \tkt{and} realized shortly after post-bounce for both progenitor models. 
Accordingly, 
\tkt{the differences in the ALP heating rates are observed to be negligible} 
before the shock revival ($\tpb\sim0.3$\,s). 
After the shock revival, the ALP heating rates show a substantial increase due to the \tkt{extension} of the gain region.

On the other hand, for the m6 model series, the cooling rates increase with a delay and are lower compared to the m2 model series. The reason for this is that it takes \tkt{a longer} time to realize a sufficient temperature to produce 600\,MeV ALPs. 
We can also see that the s25\_m6\_g08 model (dash-dot-dot red line) exhibits the higher cooling and heating rates than the s20\_m6\_g08 model (loosely dashed magenta line), because of its higher temperature. 
From our systematic investigation, it is suggested that since $Q_\mathrm{cool}$ is sensitive to the temperature, $L_\mathrm{cool}$ and $L_\mathrm{heat}$ tend to be higher in the heavier star. As a result, in the s25\_m6\_g08 model,  ALPs heat the gain region more effectively, making shock wave revival more likely than in the s20\_m6\_g08 model.

\begin{figure*}[t]
    \centering
    \includegraphics[keepaspectratio, width=17.5cm,clip]{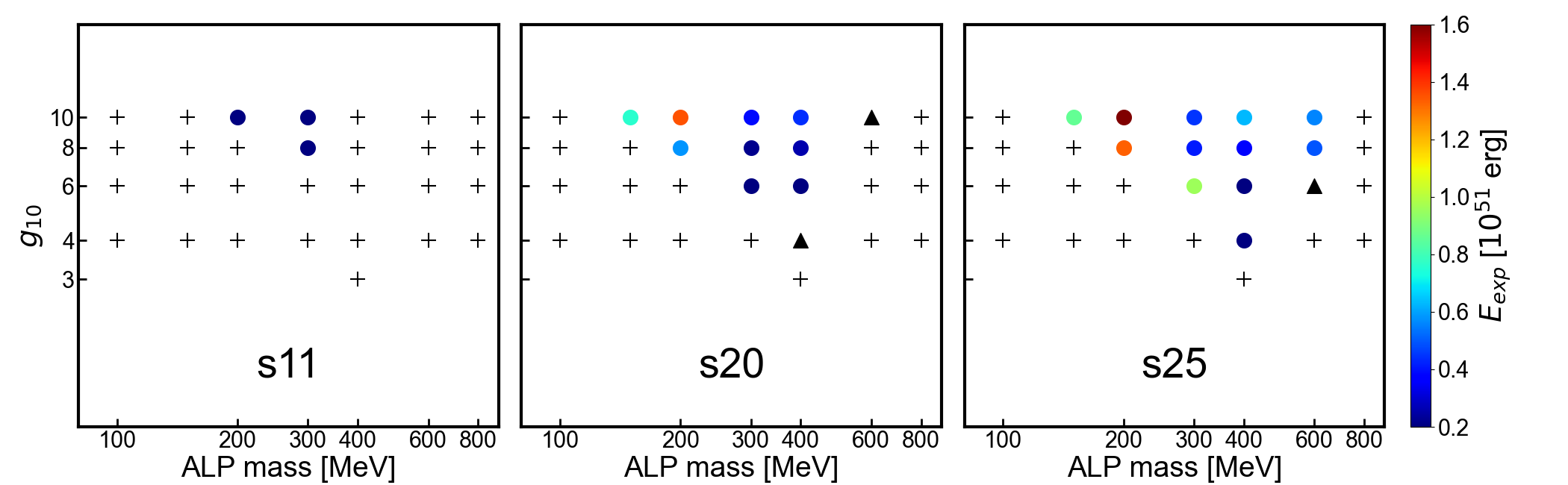}
  \caption{\label{fig:chart_progen} Outcomes of the core collapse for three progenitors with various combinations of the ALP parameters adopted in this work. The crosses represent models that fail to explode while filled circles represent models that succeed in exploding. The color shows the explosion energy when the shock wave reaches $r = 500$ km. The triangles represent models where the shock radius reaches $r= 200 -500 \km$ at $\tpb= 0.52 $\,s.
  In models incorporating ALPs, even with the same ALP parameter set, the heavier progenitors are more likely to lead to shock revival and higher explosion energy. }
\end{figure*}

In Ref.~\cite{lucente_heavy_2020}, post-processing calculations of the ALP cooling rate were performed using one-dimensional simulations that artificially amplify neutrino heating and trigger the explosion. They showed that, at $\tpb=1$\,s for the $18\,\Msun$ model, 
\tkt{the cooling rate induced by ALPs with $m_a = 600\,\MeV$}
is $2-3$ orders of magnitude lower compared to \tkt{the case for} $\ma = 200 \,\MeV$. 
However, in our study, although the progenitor model and time differ --- $20\,\Msun$ and $\tpb=0.5$\,s, respectively --- the difference in the cooling rates is only $\sim0.5$ orders of magnitude.
One of the reasons for this discrepancy is that, in our model, the delay in shock expansion for $\ma=600\,\MeV$ suppresses the decrease in mass accretion rate.  
This allows the temperature to continue increasing \tkt{up to} $6-8\,\MeV$ higher compared to \tkt{the case for} $\ma=200\,\MeV$ at $\tpb=0.5$\,s. As a result, ALPs with $\ma=600\,\MeV$ are more likely to be produced
\tkt{in our self-consistent simulations.}

Figure \ref{fig:chart_progen} shows the explodability \tkt{and explosion energies} of our SN models with various ALP parameters. 
\tkt{The symbols denote the final fate of the models and the color filling each symbol} shows the explosion energy $E_{\rm{exp}}$ at the time when the shock wave reaches $r = 500\,\km$.
In this plot, the crosses represent the models that fail to explode, while the filled circles represent the models with successful explosion. 
The triangles represent models in which the shock wave propagates outward but does not reach $r=500 \,\km$ at the end of the simulations. 
This figure indicates that the heavier progenitor models have a more extended ALP parameter region that can lead to shock revival even with the spherically symmetric geometry. It is also shown that the heavier progenitor models exhibit higher explosion energies.

\section{Conclusion and Discussion}
\label{Conclusion}

In this study, we performed stellar core-collapse simulations with ALPs to investigate the progenitor dependence of  the impact of ALPs on SNe. We employed the $11.2\,\Msun$, $20.0\,\Msun$ and $25.0\,\Msun$ progenitor models, as well as ALPs with $\ma=100-800 \,\MeV$ and $\gten=4-10$. 

\tkt{The heating effect induced by the ALP radiative decay can lead to a successful SN explosion even in the one-dimensional geometry if the heating rate $Q_\mathrm{heat}$ is sufficiently high \cite{mori_shock_2022}. Regardless of the ALP mass, when the coupling constant $g_{a\gamma}$ is higher, $Q_\mathrm{heat}$ becomes higher. When the ALP mass is lower than $\sim300$\,MeV, heavier ALPs with a fixed $g_{a\gamma}$ induce higher $Q_\mathrm{heat}$ because of a shorter mean free path of ALPs. However, when ALPs are heavier than $\sim400$\,MeV, heavier ALPs induce lower $Q_\mathrm{heat}$. This non-monotonicity is attributed to  the Boltzmann suppression for the ALP production in the SN core.}

We find that the ALP parameter region in which ALPs cause SN explosions  with the one-dimensional geometry is more extended for heavier progenitor models. Also, our simulations indicate that the explosion energy becomes higher for heavier stars. 
\tkt{For example,} for ALPs with $\ma=600\,\MeV$ and $\gten=8$, the s25 model successfully revives the shock wave, whereas the s11 and s20 models do not. This difference arises because the higher temperature environment in the heavier progenitor facilitates the production of  heavy ALPs. 

\tkt{We will refer to the limitation of this study.}
In our simulations,  the gravitational trapping effects on ALPs are not considered.
As pointed out by Ref.~\cite{lucente_constraining_2022},  ALPs can be gravitationally trapped when  the ALP kinetic energy is not large enough to escape from the stellar gravitational potential.
Although the gravitational trapping is negligible for $\ma \lesssim 100\,\MeV$ \cite{lucente_constraining_2022}, it is worthwhile to develop an advanced transport scheme to study the effect of the trapping on the ALP heating.

In addition, the ALP luminosity could be influenced by 
\tkt{the treatment of the nuclear equation of state (EoS).}
We employed \tkt{the EoS from} Lattimer and Swesty with $K=220$\,MeV \cite{lattimer_generalized_1991}, which is classified as a soft EoS.
\tkt{However, recent mass-radius measurements of neutron stars \cite{Miller_2021, Riley_2021,2021ApJ...918L..29R} 
indicate that the radius of a $1.4M_\odot$ neutron star is $\sim11-13$\,km, which is smaller than the prediction of Ref.~\cite{lattimer_generalized_1991}. From this perspective, stiffer EoSs such as the DD2F-SF \cite{PhysRevC.81.015803} and SFHo \cite{Steiner_2013} are preferred. The stiffer EoSs would reduce the PNS temperature and decrease the ALP luminosity. It remains still to quantitatively investigate  the effect of different EoSs on the ALP emission. }

\tkt{In our simulations, we focused on ALPs that interact only with photons. However, ALPs  can interact with nucleons and pions as well \cite[e.g.][]{manzari2024supernovaaxionsconvertgammarays,benabou2024timedelayedgammaraysignaturesheavy}. When the ALP-nucleon interaction is considered, ALPs can be produced through the nucleon bremsstrahlung \cite{brinkmann_numerical_1988,1995PhRvD..52.1780R,carenza_improved_2019} and the pion conversion \cite{2021PhRvL.126g1102C,fischer_observable_2021}. These processes would enhance the ALP production rate and  cause the PNS contraction because of the additional cooling \cite{betranhandy_impact_2020}. If the ALP luminosity is high enough, this effect can significantly enhance the explosion energy. In addition, the radiative decay of ALPs produced through these processes would contribute to the heating rate in the gain region. Because these effects would affect neutrino signals \cite{fischer_probing_2016,fischer_observable_2021}, detailed comparison with SN~1987A neutrinos would provide information on ALPs. In order to establish a comprehensive understanding of the role of ALPs in the SN explosion mechanism, it is desirable to extend our framework to include the ALP interactions with nucleons and pions, in addition to photons, in future studies. }

Our simulations employ spherically-symmetric geometry to perform systematic investigation across a broad range of the ALP parameter space.
However, multidimensional effects and stellar rotation play important roles on the explodability and the explosion energy \cite{janka_explosion_2012,kotake_multimessengers_2012,burrows_colloquium_2013,foglizzo_explosion_2015, nordhaus_dimension_2010,hanke_is_2012,dolence_dimensional_2013,takiwaki_comparison_2014,nakamura_long-term_2019,nagakura_towards_2019,melson_resolution_2020, foglizzo_instability_2007,blondin_stability_2003}. 
Indeed, two-dimensional SN simulations taking into account ALPs have been performed to investigate their effects on the explosion energy and neutrino and gravitational wave signals \cite{mori_multimessenger_2023}. 
Additionally, multi-dimensional models for the core-collapse of rotating stars \cite{Nakamura_2014,summa_rotation_2018} indicate that the centrifugal force could reduce the PNS temperature. Since the ALP production rate is sensitive to the matter temperature, it is desirable to develop SN models taking both stellar rotation and ALPs into account to explore the ALP effects on rotating stars.
Furthermore, three-dimensional simulations that account for ALPs have not yet been realized. 
The current study serves as a preparatory step toward developing three-dimensional simulations that incorporate ALPs, providing comprehensive insights into the important role of progenitor structure in ALPs' contribution to CCSNe. These three-dimensional simulation results are expected to offer crucial insights for constraining ALP parameters using multi-messenger signals from the next nearby supernova event.

\begin{acknowledgments} 
Numerical computations were carried out on the PC cluster at the Center for Computational Astrophysics, National Astronomical Observatory of Japan. This work is supported by JSPS KAKENHI Grant Numbers JP23KJ2147, JP23K13107, JP23K20862, JP23K22494, JP24K00631 and funding from Fukuoka University (Grant No.GR2302) and also by MEXT as “Program for Promoting researches on the Supercomputer Fugaku” (Structure and Evolution of the Universe Unraveled by Fusion of Simulation and AI; Grant Number JPMXP1020230406) and JICFuS.
\end{acknowledgments}

\appendix
\section{Analytic Expression for the ALP Heating Rate}

In our simulations, we solve Eq.~(\ref{eq:alpevol2})  to obtain the ALP heating rate $Q_{\rm{heat}}(r)$ per unit mass. Although  the numerical integration of Eq.~(\ref{eq:alpevol2}) is necessary to obtain its accurate solution, we can find an approximate expression for $Q_{\rm{heat}}(r)$ as follows. The ALP luminosity is written as 
\begin{equation}
    L_{a}(r) = 4\pi r^{2} \mathcal{F} (r) = L_{0} \exp\left(- \frac{r - r_\mathrm{max}}{\lambda}\right),
\end{equation}
where $\lambda$ is the mean free path of ALPs, $\mathcal{F}(r)$ is ALP flux, $r_\mathrm{max}\sim10$\,km is the peak radius of $Q_\mathrm{cool}$, and 
\begin{equation}
    L_{0} = 4\pi \int dr r^{2} Q_\mathrm{cool} (r).
\end{equation}
The energy per second injected by the ALP into the spherical shell of $r$ to $r +\Delta r$, $L_\mathrm{heat}(r)$, is
\begin{align}
    L_{\rm{heat}}(r) &= L_{a}(r) - L_{a}(r + \Delta r) \nonumber\\
    &= L_{0}\left(\exp\left(-\frac{r-r_{\rm{max}}}{\lambda}\right)-\exp\left(-\frac{r+\Delta r - r_{\rm{max}}}{\lambda}\right)\right)\nonumber \\
    &= L_{0} \exp\left(-\frac{r - r_{\rm{max}}}{\lambda}\right)\left(1-\exp\left(-\frac{\Delta r}{\lambda}\right)\right)\nonumber \\
    &\approx L_{0} \exp\left(- \frac{r-r_{\rm{max}}}{\lambda}\right) \frac{\Delta r}{\lambda}.
\end{align}
The heating rate $Q_{\rm{heat}}(r)$ can be evaluated as
\begin{equation}
\label{eq:qheat}
    Q_{\rm{heat}} (r) = \frac{L_{\rm{heat}} (r)}{4\pi r^{2} \Delta r} \approx  \frac{L_{0}}{4\pi r^{2} \lambda} \exp\left(- \frac{r-r_{\rm{max}}}{\lambda}\right).
\end{equation}
The analytical estimation indicates that  $Q_{\rm{heat}}$ is approximately proportional to $\gten^4$ because  $Q_{\rm{cool}}$ are proportional is $\gten^2$. One can also see that, when $r-r_\mathrm{max}\ll \lambda$, $Q_\mathrm{heat}(r)$ obeys the inverse-square law.

\bibliography{takata}

\end{document}